\documentclass[sigconf,
               nonacm,
              ]{acmart}

\usepackage{hyperref}
\usepackage{enumitem}
\usepackage{color, colortbl}
\usepackage[english]{babel}
\usepackage[utf8]{inputenc}
\usepackage{algorithm}
\usepackage[noend]{algpseudocode}
\usepackage{multirow}
\usepackage{tabularx}
\usepackage{booktabs}
\usepackage{pifont}
\newcommand{\xmark}{\textcolor{red}{\ding{55}}}
\newcommand{\cmark}{\textcolor{green}{\ding{51}}}
\definecolor{Gray}{gray}{0.9}

\begin{document}

\title{ICtoken: An NFT for Hardware IP Protection}

\author{Shashank Balla}
\affiliation{%
  \institution{University of California, San Diego}
  \city{San Diego}
  \state{California}
  \country{USA}
  }
\email{sballa@ucsd.edu}

\author{Yiming Zhao}
\affiliation{%
  \institution{University of California, San Diego}
  \city{San Diego}
  \state{California}
  \country{USA}
  }
\email{yiz060@ucsd.edu}

\author{Farinaz Koushanfar}
\affiliation{%
  \institution{University of California, San Diego}
  \city{San Diego}
  \state{California}
  \country{USA}
  }
\email{farinaz@ucsd.edu}

\begin{abstract}

Protecting integrated circuits (ICs) from piracy and theft throughout their lifecycle is a persistent and complex challenge. In order to safeguard against illicit piracy attacks, this work proposes a novel framework utilizing Non-Fungible Tokens (NFTs) called ICtokens, uniquely linked to their corresponding physical ICs. Each ICtoken contains comprehensive information, including authentication data, supply chain stage and status, ownership details, and other IC metadata, while also making provision for the secure integration of a logic-locking key. Designed to be publicly logged, ICtokens securely obscure metering information without compromising functionality. In addition, the ICtracker, a distributed ledger technology powered by a swift and energy-efficient consortium blockchain, is used to register and manage ICtokens and their respective owners, tracking all associated interactions. This robust ledger guarantees the traceability and auditing of ICtokens while simultaneously developing a product-level NFT at every transaction point within the supply chain. Consequently, a scalable framework is established, creating unique, immutable digital twins for ICs and IC-embedded products in the form of ICtokens and their transactions. This provides a robust and reliable supply chain trail back to the original IP owner, while also offering unprecedented assurance to consumers of IC-embedded products. The rich information contained within ICtokens facilitates more detailed audits than previous proposals for IC supply chain monitoring. A proof-of-concept, implemented as an open-source solution, ensures the ease of adoption of the proposed framework.

\end{abstract}



\maketitle

\section{Introduction}

\begin{figure}[h]
  \centering
  \includegraphics[width=\linewidth]{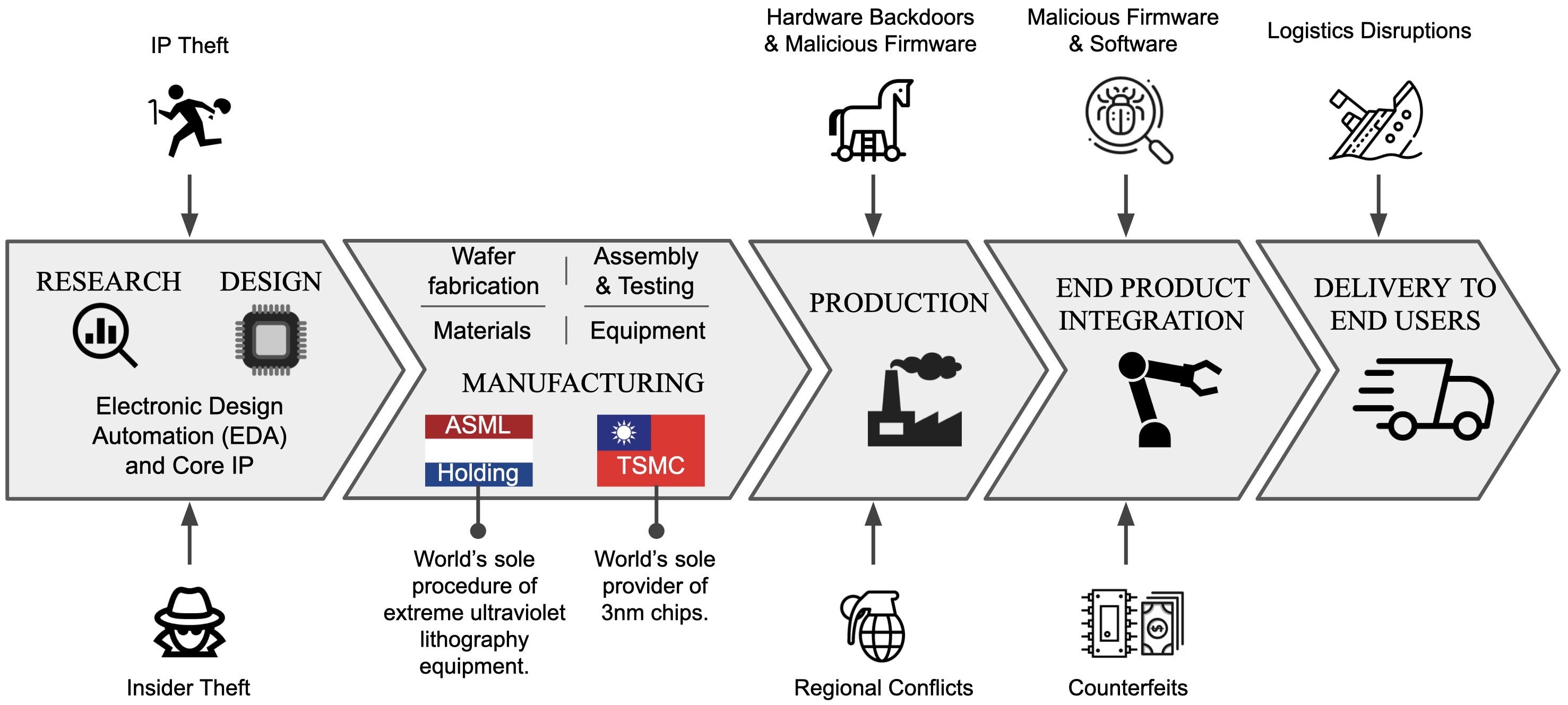}
  \caption{Supply chain threats. \cite{scsupshainrisks}}
  \label{fig:supplychain}
\end{figure}

Today, the microscopic size and negligible cost of ICs has enabled the addition of logical components to many devices, making them \emph{smart}. A plethora of new businesses  creating niche products have come up that require rapid prototyping of their products. To cater to this incredible demand the semiconductor industry has adopted a horizontal supply chain model. During the lifetime of an IC, multiple parties take up distinct roles;
designers outsourcing fabrication, distributors stocking up/reselling, assemblers/integrators sourcing components, and original equipment manufacturers building the end products. Unfortunately, this highly distributed setting has also opened up many avenues for IP theft \cite{scsupshainrisks} [Figure:\ref{fig:supplychain}]. Recently, the global shortage of chips has exacerbated the issue, where bootleggers pirate name-brand designs, claim high-quality assurances \cite{Newton22}. The infiltration of counterfeit products into the supply chain puts end-consumers and critical infrastructure at high risk \cite{usdoc}, \cite{goa16}.

Many techniques have been proposed by the research community to tackle the various threats of hardware piracy \cite{Rajendran17}. Rostami et al \cite{Rostami14} and Guin et al \cite{guin2014detectandprevent} survey distinct attacks at each stage of the supply chain for various threat models. For each of these attacks, they discuss methods for detecting it as well as techniques to prevent such an attack. The main limitation of these physical detection methods is that the application of just one is not sufficient to detect all attacks. Further, most of them destroy the Device-Under-Test in addition to being very laborious and requiring very costly specialized equipment. These techniques are impractical for businesses to deploy and impossible for individuals without the necessary tools. However, some of the prevention techniques, as discussed below, have seen commercial use with IP owners protecting their IPs to a substantial extent with the foundries.

Logic locking is a pre-silicon hardware IP protection technique that obfuscates the functionality of an IC by design until a correct key (known only to the IP owner) is provided along with the input \cite{Yasin17} , \cite{kamali22}. Thus, even in fabrication, malicious actors who have physical access to the chip will not obtain the correct input-output behavior of the IC, as, at that point, only the IP owner has the key. Though this approach offers an effective IP protection mechanism against overbuilding and reverse engineering, it does not consider the post-fabrication testing of the die required for quality assurance purposes before it is packaged and released. Secure Split-Tests (SSTs) help overcome this challenge \cite{Contreras13}. SSTs augment logic locking by making provisions for testing on the locked designs. This process requires rounds of communication between the IP owner and fabrication house to test the functionality of chips. A particular die is approved for packaging only after the IP owner is satisfied with the functionality. 

For post-silicon fingerprinting and  authentication, Physical Unclonable Functions (PUFs) is  tried-and-tested technology \cite{Serge18}, \cite{intrinsic}, \cite{radicalSC}. PUFs leverage random process variations at the fabrication stage to generate unique in-borne identifiers for the IC. These identifiers can then be used for authentication and attestation of the IC by the device owners and guard against cloning \cite{herder14}. The above two approaches, Logic-locking and PUFs, can be fused together to create device-unique keys that can obfuscate the functionality of an IC \cite{Roy08}. Techniques that create device-specific unique identifiers which can also be used to obfuscate functionality of the IC are referred to as Active Metering \cite{Koushanfar10}. 

\par
Such methodologies do provide protection against piracy attacks, but they do not encompass the integrity of ICs throughout the entire supply chain. Hence, there is a need for a supply-chain-wide foolproof framework that builds a guarantee in the product as it evolves through various transactions with the full knowledge of the IP owner. In this context, developing trustworthy transactions throughout the supply chain becomes the bedrock of a secure framework for both IP protection as well as a guarantee to the end-user. For a trustworthy transaction on any product, the buyer must be able to verify the authenticity of the product and the seller must be able to ensure licensed use of IP within the product among other things. A publicly accessible database incorporating relevant metadata of ICs as updated through transactions would help to authenticate registered products and detect any discrepancies. Blockchains are a popular approach to creating immutable public distributed ledgers and have been widely adopted by many businesses to offer supply chain monitoring services \cite{Nehra21}, \cite{EY22}, \cite{skuchain}.

\begin{figure}[h]
  \centering
  \includegraphics[width=\linewidth]{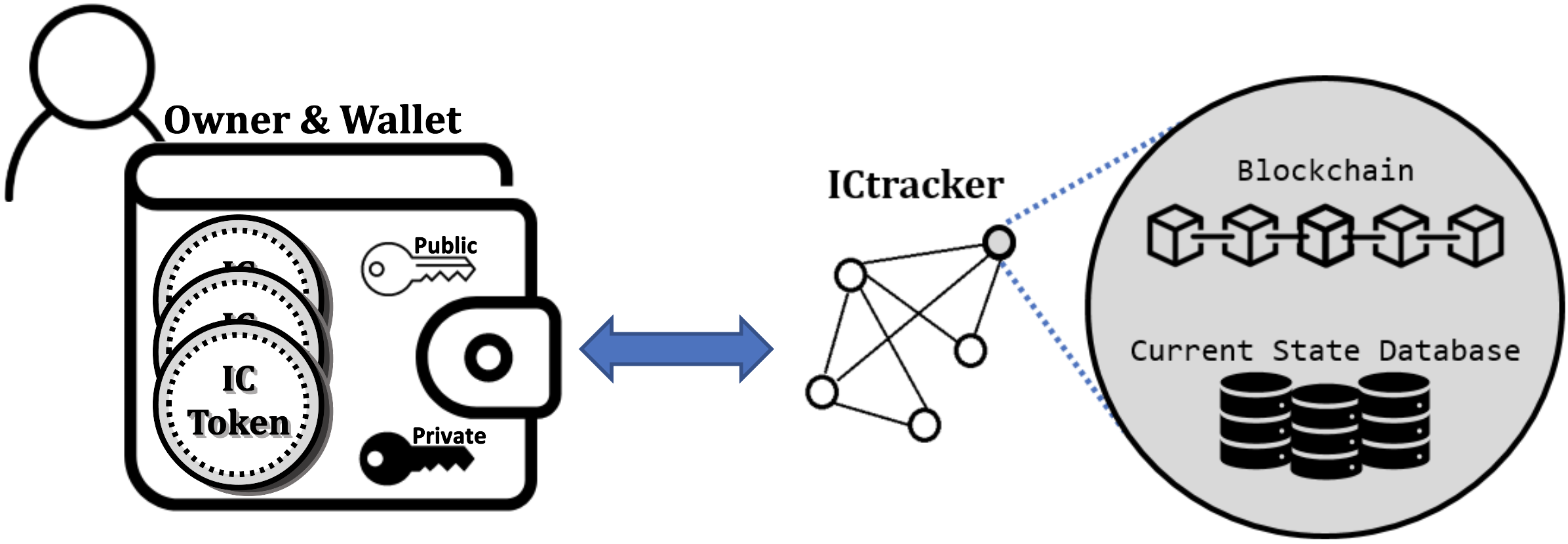}
  \caption{Components of the proposed framework.}
  \label{fig:ICtracker}
\end{figure}

This paper suggests ICtoken, a novel framework (Figure \ref{fig:ICtracker}) to enable provable owner tracking of the ICs through the long supply chain. The contributions are as follows: 
\begin{itemize}
  \item Introduction of ICtoken, assigning an NFT per IC. The individual NFTs can be leveraged to create product-level digital assets.
  \item Provisions for secure transfer of IC specific information (e.g. active metering information) from one owner onto the next while ensuring privacy and tamper-resistance.
  
  \item Introduction of ICtracker, an essential block-chain component of the framework which can be utilized to trace ICs through its product life cycle.

  \item The robustness of ICtoken framework is demonstrated through analysis of the block-chain based method against various contemporary threats.
\end{itemize}

We review the related background underlying our motivations in Section 2. In Section 3, we describe in depth our framework with ICtokens and the ICtracker. Finally, we evaluate our framework by showing its resistance to various IC supply chain threats in Section 4.

\section{Background}

Blockchain technology is a path breaking approach of recent times with revolutionary applications in several industries \cite{Nehra21}, \cite{walmartFood}, \cite{fedex}. Blockchain is an immutable decentralized hash chain that stores information in a linked list of blocks which is mirrored on all nodes participating in the network.  Every block is linked to the previous one by carrying a cryptographic hash of all the data stored in the previous one. Information is stored on the blockchain in its header field and a consensus protocol ensures that this information is mirrored across all nodes in the network. Bitcoin \cite{Nakamoto08} was the first application to popularize the use of blockchains to log transactions on digital currency. Later applications also allowed more types of information to be stored on this database. Ethereum \cite{Buterin14} introduced smart contracts which are Turing complete programs that can be stored on the blockchain, immutably binding the contracted parties to the terms of the contract. Four years ago, the introduction of NFTs added further impetus to the use of blockchains. They enable users to digitize their assets by pushing a digital signature of it onto a blockchain thereby establishing the signing entity’s ownership \cite{Wang21}. Ethereum first proposed the design of NFTs in an Ethereum Improvement Proposal (EIP) \cite{eip721}. This proposal introduced the ERC 721 standard \cite{erc721} to define the interface and functionalities a smart contract must employ to be able to mint NFTs. This effectively created a marketplace for digital assets to be traded on a blockchain where original creators could also get paid royalties every time their NFT was traded \cite{opensea}.

One of the initial works that considers blockchains for IC supply chain monitoring is Islam et al \cite{Islam19}. They propose to use a smart contract deployed on a blockchain which has provisions for enrolling an IC, authenticating it and transferring its ownership. The information they log for every IC is a unique identifier, its current owner and PUF-derived challenge and response pairs (CRPs). The storage of CRPs on the blockchain, including it in every transaction, increases the size posing scalability issues. Chaudhary et al \cite{Chaudhary21} aim to mitigate this issue by exporting the CRPs to an off-chain system called InterPlanetary File System (IPFS) \cite{Benet14} that exists as a storage oracle for the blockchain to access and retrieve files. However, \cite{Islam19} and \cite{Chaudhary21} consider only one stage of production, that is, system integration, and do not cover the prior stages of production.

Cui et al \cite{Cui19} propose a two-step transaction process to eliminate errors that cause asset transfer to a wrong address and inadvertently risk losing proprietorship of the asset, namely the IC. But they do not track the production stages or the work that was done on the IC while it was in the hands of each of its owners. Xu et al \cite{Xu19}, Vosatka et al \cite{Vosatka20} propose to start tracking right after fabrication and assume active metering techniques for protection against fabrication level counterfeiting attempts. However, they make no provision for saving this information at the blockchain requiring all future owners to interact with the IP owner to retrieve this.  

Concurrently, there has been progress in the domain of hardware implementations for improved IP integrity for IC designers. Logic locking is a prominent technology, where the latest advances give cryptographic security guarantees. Beerel et al \cite{Beerel22} lay down mathematical security definitions for logic locking. Chhotaray et al \cite{Chhotaray21} and Ganji et al \cite{Ganji19} propose cryptographically secure logic locking schemes. Further, \cite{Ganji19} leverage blockchain technology to register gate-level representations of an IC as IP blocks and enable pay-per-use evaluations. Specifically, they allow limited number of accesses to the IC while all previous schemes lose the effectiveness of the logic locking key upon first access of the IC.

The proposed framework addresses the shortcomings of previous works while integrating the latest developments to strengthen IP integrity for IC designers across the product life cycle. It enables registration, authentication, active metering and end-to-end tracing in a privacy-preserving fashion, duly implemented on a transparent blockchain base.

\section{Framework Design}

We model the complete IC supply chain with three distinct constituent parts. ICtokens represent the merchandise; ICs, Printed Circuit Boards (PCBs) and Electronic Devices. Owners are the transactors of this merchandise and therein the ICtokens. Lastly, ICtracker is the encompassing protocol and moderation system for all transactions.

\subsection{ICtoken}

Here we introduce the design template for constructing an ICtoken i.e. an NFT uniquely bound to a physical IC capturing all movements throughout its life cycle. The metadata of the ICtoken has all the necessary information to achieve this. To ensure binding, we must tag ICtoken with the same identifier as the physical IC; and for this purpose we include \textit{ICID}, a SHA256 hash of the unique identifier for the IC. Further, as the IC progresses through the supply chain, it might get built into a PCB, which can then be uniquely identified by the set of ICs embedded in it. A unique identifier for the PCB, a \textit{PID}, can be constructed from a merkle hash of the \textit{ICIDs}. To tag the ICtoken to its PCB, the \textit{PID} is included in its metadata as well. Similarly, when this PCB gets built into an electronic device (system), it can further be uniquely identified by the set of PCBs used in it and an \textit{EDID} can be constructed from a merkle hash of their \textit{PIDs}. To tag the ICtoken with the electronic device, the \textit{EDID} is also stored in the metadata. To capture the markings on the IC's package, a SHA256 hash of its markings, \textit{markHash}, is included. The production \textit{stage} (1: Fabrication; 2: PCB Assembly; 3: System Integration; 4: End-User) and \textit{status} (0: In progress; 1: Completed) of the IC are included to track its progression in the supply chain. For provenance, \textit{prevVer}, a linker to the previous version of the ICtoken, and \textit{version}, the current version number of the ICtoken are also included. Finally, a flag, \textit{isDefective}, is present to notify if the IC has been reported to be malfunctioning.

\begin{table}[h]
    \centering
    \caption{ICtoken Size Analysis}
    \begin{tabular}{|l|l|r|}
        \hline
        \textbf{Attributes}&\textbf{Data}&\textbf{Size} \\
        \hline
        Metadata&\textit{ICID, PID, EDID}  & 3 $\times$ 32B\\ 
                &\textit{markHash}        & 32B            \\
                &\textit{Stage, Status}   & 3b,  1b        \\
                &\textit{prevVer, version}& 8B, 1B         \\
                &\textit{isDefective}     & 1b             \\
        \hline
        ICkey   &\textit{keyEncr}         & 256B           \\
                &\textit{keyHash}         & 32B            \\
        \hline
        Owner   &\textit{publicID}        & 32B            \\
        \hline
        \textit{trnsaxnID}&\multicolumn{2}{r|}{256B}\\
        \hline
        \textbf{Total}
        &\multicolumn{2}{r|}{\textbf{714B}}\\
        \hline
    \end{tabular}
    \label{tab:ictoken}
\end{table}

By design, ICtoken also has provisions to securely carry the \textit{key} for active metering. The ICkey attribute holds an encryption of the \textit{key} under the current owner’s public key, \textit{keyEncr}, and an SHA256 hash of the \textit{key}, \textit{keyHash}, to verify correct decryption as well as to ensure a consistent trail with previous versions. More generally , the ICkey attribute is provisioned for in the template to securely store any information even if active metering is not employed. Proprietorship information is tracked through the owner attribute which carries the \textit{publicID} of the current owner. And lastly, the \textit{transaxnID} attribute is the owner’s digital signature on all the above information, binding it as the source. Table \ref{tab:ictoken} analyses the total size of an ICtoken object.

\subsection{Owner}

An Owner represents a participant in the IC supply chain, and can be any entity that possesses or wishes to possess an IC at any stage in its life cycle. These prospective owners could be hardware designers (IP holders), fabrication units, PCB assemblers, system integrators, end users, or electronics recyclers, coming in at different production/use stages of an IC, as well as any intermediate distributors/resellers. Thus, the variety of possible owners considered is comprehensive, which ensures every IC explicitly has an owner at any instance in its life-cycle. Every owner requires a unique \textit{publicID} and a \textit{public-private key pair} to participate in the framework's protocol. The \textit{publicID} is used for identification and the \textit{public-private key pair} is used towards establishing secure communication and transactions. The \textit{public-private key pair} enables the following functionalities for the owner:

\begin{itemize}
    \item \texttt{encrypt} and \texttt{decrypt}: Encrypts/decrypts the input with the \textit{public}/\textit{private key}.
    \item \texttt{changeEncKey}: Changes the encryption key of the input to a \textit{public key} of another owner; decrypt input and encrypt result with new owner's \textit{public key}.
    \item \texttt{signMessage}: Creates a digital signature of a message with the \textit{private key}.
    \item \texttt{verifySign}: Verifies a digital signature with the \textit{public key}.
\end{itemize}

Every Owner has a \textit{wallet} that authorizes its interface with the blockchain and stores all owned ICtokens. The \textit{wallet} stores the \textit{publicID}, the \textit{public-private key pair} of the owner, and a public profile of the owner consisting of the \textit{publicID}, \textit{public key} attributes and the \texttt{verifySign} functionality. The \textit{wallet} uses the public profile to enroll with the blockchain and authorize transactions to/from other wallets on the blockchain.

\subsection{ICtracker}

ICtracker is a blockchain-based public distributed ledger that keeps track of enrolled ICs throughout their product life cycle by logging all transactions of ICtokens. It maintains a database of the current state of the blockchain, public profiles of all enrolled owners and the ICtokens they currently own. ICtokens are logged as transactions in our blockchain. Since every block is of limited size it can store only a fixed number of ICtokens.
ICtracker receives transactions from a wallet in the form of new ICtokens while requesting for a particular service. ICtracker will perform the necessary checks to ensure that the new ICtokens have only those changes that are permitted by the service requested. If these requirements are met, ICtracker adds the new ICtokens to the latest block and updates its current state database when the addition is committed. In the distributed setting, this protocol will run on every node in the network, and any transactions of ICtokens will require a consensus across all nodes in the network. Every node stores a local copy of the latest committed state of the blockchain along with four mappings in its database that keeps up with this state. These mappings are paramount for authentication and verification of transactions and overall efficiency of the protocol. The mappings are as follows: 

\begin{description} 
\item [\texttt{ICdb }:] \textit{ICID} → latest index in the blockchain.
\item [\texttt{PCBdb}:] \textit{PID} → list of \textit{ICIDs} 
\item [\texttt{DEVdb}:] \textit{EDID} → list of \textit{PIDs} 
\item [\texttt{OWNdb}:] \textit{publicID} → \{\textit{isEnrolled, pubProf, assets}\}
\end{description} 

ICdb quickly identifies the index of a particular \textit{ICID} on the blockchain. Specifically, it retrieves the latest version of the ICtoken for an \textit{ICID} on the blockchain. It is useful for auditing, i.e.,  when someone wishes to verify the most up-to-date information for a particular IC. Moreover, since the ICtoken also stores the index of its previous version, it is simple to trace back the entire history of the IC. PCBdb and DEVdb keep track of the PCB-level and System-level tokens, i.e., collections of the ICtokens used in the device. OWNdb stores the information of enrolled owners, their public profile and all assets (ICs, PCBs and Devices) currently associated with their account. It makes for quick verification and authentication of owners on the node’s end and faster updates for owners’ wallets to find all the latest ICtokens linked to their accounts. 

ICtracker provides the following services for owners to transact on ICtokens.
\begin{itemize}
    \item \texttt{enrollOwner}: Lets owners enroll themselves as authorized users by storing their data in OWNdb.  
    \item \texttt{verifyTransaxn}: Verifies the authenticity of the transaction by validating the owner's signature in the ICtoken's \textit{trnsaxnID} with the owner's \textit{public key}. It aborts the service requested if verification fails.
    \item \texttt{enrollIC}: Lets enrolled owners register new ICs into ICtracker for the first time. [Algorithm \ref{alg:ICtracker_enrollIC}]
    \item \texttt{reportDefective}: Lets users report malfunctioning ICs. ICtracker verifies the transaction and sets the \textit{isDefective} bit of the ICtoken if successful.
    \item \texttt{updateStage}: Lets owners update the \textit{stage} or \textit{status} of a single ICtoken. [Algorithm \ref{alg:ICtracker_updateStage}]
    \item \texttt{updatePIDorEDID}: Lets owners update the \textit{PID} or \textit{EDID} of multiple ICtokens. [Algorithm \ref{alg:ICtracker_updatePIDorEDID}]
    \item \texttt{transferIC}: Lets owners transfer their ICtoken to the next owner. [Algorithm \ref{alg:ICtracker_updatePIDorEDID}]
\end{itemize}

\begin{figure}[h]
  \centering
  \includegraphics[width=\linewidth]{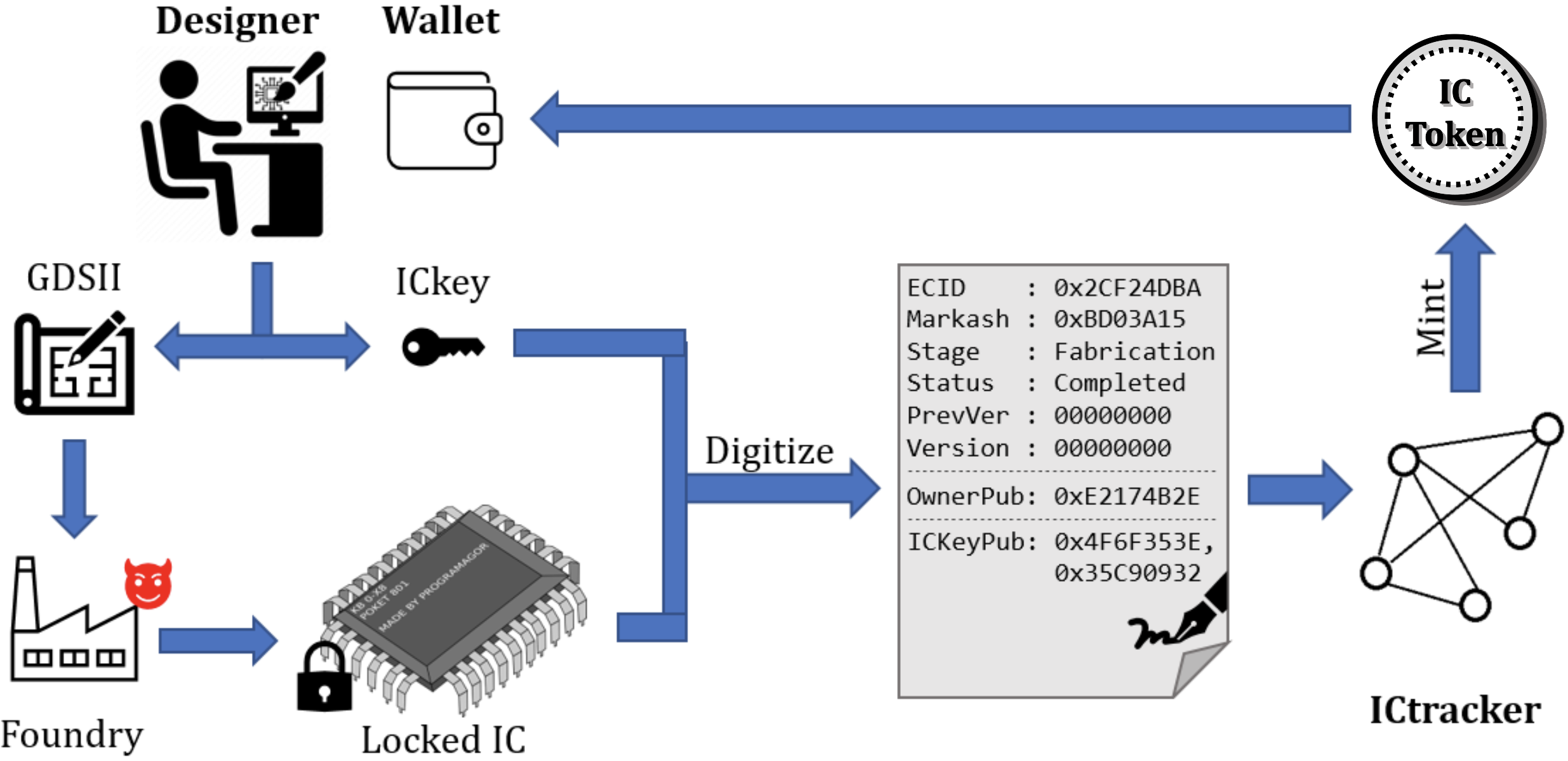}
  \caption{IC enrollment.}
  \label{fig:ICenrollment}
\end{figure}

\begin{algorithm}
\caption{Enrolling an IC in ICtracker}
\label{alg:ICtracker_enrollIC}
\begin{algorithmic}[1]
\Procedure{\texttt{enrollIC}}{ICtoken}:
\State set ICID = ICtoken.metaData.ICID
\State \textbf{assert} ICdb[ICID] == None \Comment can't re-enroll
\State set ownID = ICtoken.owner.publicID
\State verifyTransaxn(OWNdb[ownID], ICtoken)
\State \textbf{assert} ICtoken.metaData.stage  == \textit{Fabrication} 
\State \textbf{assert} ICtoken.metaData.status == \textit{Completed} 
\State \textbf{assert} ICtoken.metaData.PID == None
\State \textbf{assert} ICtoken.metaData.EDID == None
\State \textbf{assert} ICtoken.metaData.prevIdx == None
\State \textbf{assert} ICtoken.metaData.version == 0
\State add ICtoken to blockchain \Comment{Data matches a new IC}
\State update ICdb, OWNdb with latest ICtokens and assets.
\EndProcedure
\end{algorithmic}
\end{algorithm}

\begin{algorithm}
\caption{Updating an IC's stage/status in ICtracker}
\label{alg:ICtracker_updateStage}
\begin{algorithmic}[1]
\Procedure{\texttt{updateStage}}{ICtoken}:
\State \textbf{assert} ICID of ICtoken is enrolled and not defective.
\State set metaData = ICtoken.metaData
\State fetch prevToken from ICdb
\State verifyTransaxn(OWNdb[prevToken.owner], ICtoken)
\State set prevData = prevToken.metaData \Comment{previous metadata}
\State \textbf{assert} metaData == prevData other than stage, status
\State \textbf{assert} metaData.stage  >= prevData.stage
\If{metaData.status < prevData.status}
\State \textbf{assert} metaData.stage  > prevData.stage
\EndIf
\State ICtoken.version += 1 \Comment{increment version}
\State set ICtoken.prevVer = ICdb[ICID] \Comment{add link to previous}
\State add ICtoken to ICtracker.blockchain
\State update ICdb, OWNdb with latest ICtokens and assets.
\EndProcedure
\end{algorithmic}
\end{algorithm}

\begin{algorithm}
\caption{Updating PID/EDID of multiple ICs in ICtracker}
\label{alg:ICtracker_updatePIDorEDID}
\begin{algorithmic}[1]
\Procedure{\texttt{updatePIDorEDID}}{ICtokens}:
\State \textbf{assert} all ICtokens' ICIDs are enrolled and not defective.
\State fetch prevTokens of all ICtokens
\For{every ICtoken in ICtokens} \Comment{check all owners same}
    \State verifyTransaxn(previous owner, ICtoken)
\EndFor
\State \textbf{assert} ICtokens == prevTokens other than PID, EDID.
\State \textbf{assert} prevTokens' status is \textit{Completed}.
\If {all ICtokens have same PID}       \Comment{update PID}
    \State \textbf{assert} all ICtokens have stage == \textit{PCB Assm}
    \State \textbf{assert} all ICtokens have EDID == None
    \State \textbf{assert} all prevTokens have PID and EDID == None
    \State PCBdb[PID] = list(ICIDs) \Comment update PCBdb
\ElsIf{ all ICtokens have same EDID}    \Comment{update EDID}
    \State \textbf{assert} PID of all ICtokens, prevTokens is same
    \State \textbf{assert} all ICtokens have stage == \textit{Sys Int}
    \State \textbf{assert} all prevTokens have EDID == None
    \State DEVdb[EDID] = list(PIDs) \Comment update DEVdb
\EndIf 
\State \textbf{else} EXIT \Comment{invalid data}
\For{every ICtoken in ICtokens}
    \State ICtoken.version += 1     \Comment{increment version}
    \State add ICtoken to ICtracker.blockchain
    \State update ICdb, OWNdb with latest ICtokens and assets.
\EndFor
\EndProcedure
\end{algorithmic}
\end{algorithm}

\begin{algorithm}
\caption{Transferring IC ownership in ICtracker}
\label{alg:ICtracker_transferIC}
\begin{algorithmic}[1]
\Procedure{\texttt{transferIC}}{ICtoken}:
\State \textbf{assert} ICID of ICtoken is enrolled and not defective.
\State \textbf{assert} ICtoken.owner is enrolled \Comment{new owner}
\State fetch prevToken for the ICID
\State \textbf{assert} prevToken's status is \textit{Completed}.
\State \textit{verifyTransaxn}(prevToken.owner, ICtoken)
\State \textbf{assert} ICtoken == prevToken other than key, owner
\State ICtoken.version += 1     \Comment{increment version}
\State add ICtoken to ICtracker.blockchain
\State update ICdb, OWNdb with latest ICtokens and assets.
\EndProcedure
\end{algorithmic}
\end{algorithm}

\section{Protocol analysis}

\subsection{Feasibility analysis}
We have made two open-source implementations for the framework proposed in the paper \cite{implementations}. The first one is an interactive python notebook where all components of the framework i.e. ICtokens, Owners, Wallets, ICtracker, and all their functionalities as proposed are codified. The python notebook simulates the complete protocol for the interface between owners’ wallets and a node in the ICtracker network. Thus, it serves as an accurate blueprint for establishing the necessary blockchain infrastructure required for the framework from the ground up. The second implementation is targeted towards deploying the framework on pre-existing infrastructure. Here we developed a smart contract for ICtracker in solidity programming language \cite{solidity} which is supported by various prominent blockchain platforms \cite{Buterin14}, \cite{monax}. We compile and deploy this as an application on the Ethereum blockchain. Once deployed, this smart contract can mint ICtokens, as presented in the paper, that are compliant with the ERC721 standard \cite{erc721}. It also enables similar transactions on ICtokens by providing all the services that were proposed to be offered by ICtracker in the paper.  

Our implementations serve as a concrete proof-of-concept for the proposed framework by showing all the necessary functionalities and infrastructure required to solve the problem of hardware IP protection in a globalized supply chain and its feasibility of providing these functionalities through implementation on a pre-existing general-purpose public blockchain.

\begin{table*}[h!]
    \centering
    \begin{tabular}{|c|l|c|c|c|c|l|}
    \hline
         \textbf{Version}&\multicolumn{1}{|c|}{\textbf{Function}}&\textbf{Owner}&\textbf{Stage/ Status}&\textbf{PID}&\textbf{EDID}& \multicolumn{1}{|c|}{\textbf{Description}}  \\
         \hline
         \rowcolor{Gray}
         \multicolumn{7}{|c|}{GDSII → IC}\\
         \hline
         1&\texttt{enollIC()}&Owner1&1/ 1& Null & Null& Owner1 enrolls the IC after fabrication.\\
         \hline
         2&\texttt{transferIC()}&\cellcolor{green!25}Owner2&1/ 1& Null & Null& Owner1 transfers the IC to Owner2\\
         \hline
         3&\texttt{updateStage()}&Owner2&\cellcolor{green!25}2/ 0& Null & Null& Owner2 takes IC to PCB assembly\\
         \hline
         \rowcolor{Gray}
         \multicolumn{7}{|c|}{IC → PCB}\\
         \hline
         4&\texttt{updateStage()}&Owner2&\cellcolor{green!25}2/ 1& Null & Null& PCB assembly finished\\
         \hline
         5&\texttt{updatePIDorEDID()}&Owner2&2/ 1& \cellcolor{green!25}"......" & Null& Owner2 updates PID for all ICs in PCB\\
         \hline
         6&\texttt{transferIC()}&\cellcolor{green!25}Owner3&2/ 1& "......" & Null& Owner2 sells the PCB to Owner3 (distributor)\\
         \hline
         7&\texttt{transferIC()}&\cellcolor{green!25}Owner4&2/ 1& "......" & Null& Owner3 sells the PCB to Owner4\\
         \hline
         8&\texttt{updateStage()}&Owner4&\cellcolor{green!25}3/ 0& "......" & Null& Owner4 takes PCB to System Integration\\
         \hline
         \rowcolor{Gray}
         \multicolumn{7}{|c|}{PCB → Device}\\
         \hline
         9&\texttt{updateStage()}&Owner4&\cellcolor{green!25}3/ 1& "......" & Null& System Integration finished\\
         \hline
         10&\texttt{updatePIDorEDID()}&Owner4&3/ 1& "......" &\cellcolor{green!25}"......" & Owner4 updates EDID for all ICs in SoC\\
         \hline
         11&\texttt{transferIC()}&\cellcolor{green!25}Owner5&3/ 1& "......" &"......" & Owner4 sells the SoC to Owner5 (retailer)\\
         \hline
         12&\texttt{transferIC()}&\cellcolor{green!25}Owner6&3/ 1& "......" &"......" & Owner5 sells the SoC to Owner6 (end user)\\
         \hline
         13&\texttt{updateStage()}&Owner6&\cellcolor{green!25}4/ 0& "......" &"......" & Owner6 updates the stage during device setup\\
         \hline
         14&\texttt{updateStage()}&Owner6&\cellcolor{green!25}4/ 1&"......" &"......" & Owner6 updates the stage before recycling\\
         \hline
         15&\texttt{transferIC()}&\cellcolor{green!25}Owner7&4/ 1& "......" &"......" & Owner6 sells the device to Owner7 (recycler)\\
         \hline
         \rowcolor{Gray}
         \multicolumn{7}{|c|}{End of Life}\\
        \hline
    \end{tabular}
    \caption{Life cycle of an ICtoken from design to recycling}
    \label{tab:Tracking_framework}
\end{table*}

\subsection{Reliability analysis}

Here we analyse how our protocol ensures reliability by incorporating the following design principles in all of its functionalities.

\subsubsection{Authenticity}
Any service performed by ICtracker first begins with verifying the authenticity of the transaction. This is done by verifying the ICtoken's transaction ID, \textit{trnsaxnID}, which is a digital signature of the owner on all the data stored in the ICtoken. It is verified with the Owner’s \textit{public key} that is stored in ICtracker’s database. This ensures that only authenticated and registered users/ owners of the ICtoken are permissioned to change any data stored on the ICtoken. Furthermore, an ICtoken can only be transferred to an enrolled owner.
\subsubsection{Access control}
The active metering information for the IC is very confidential information. This must be accessible only to the current owner. To ensure this, ICtokens by design only store an encryption of this information under the current owner's \textit{public key}. Furthermore, the presence of a cryptographic hash of the metering information ensures that it has not been tampered with.

\subsubsection{Component privacy}
For the physical IC, all its metadata is stored as a cryptographic hash in its ICtoken. Pre-image resistance of the hash function employed guarantees that the original information cannot be duplicated or spoofed simply by viewing the contents on the ICtoken. Furthermore, it also ensures that any entity even with full access to all information stored in ICtracker cannot learn anything about ICs that are not in its possession.

\subsubsection{Device Integrity}
The framework also generates device level tokens as a composition of the basic ICtokens built into it. The identifiers for these devices are generated from a merkle hash of all the ICtokens integrated into it. This ensures swapping one component with another will be impossible without causing a change to the device identifier. Moreover, once a device identifier is loaded into the ICtoken for the first time, it can never be modified. This ensures that reuse of ICs is not possible.

\subsubsection{Production Integrity}
For any of the electronic device production stages, the \textit{stage} or \textit{status} attributes of an ICtoken can only be incremented [Algorithm \ref{alg:ICtracker_updateStage}]. This offers rollback protection, ensuring that ICs in the production phase do not re-enter the supply chain at earlier stages. Moreover, ICtracker allows adding device identifiers to ICtokens only at the relevant stages of production. This ensures a one-way progression of an IC through the production process. In table \ref{tab:Tracking_framework} we analyse our complete protocol, by following the ICtoken of an IC as it progresses from fabrication to end of life.

\subsection{Security}

We now evaluate our proposed framework on the various threats in an untrusted supply chain.

\subsubsection{Overbuilding and Cloning}
Overbuilding or Cloning of chips is a threat made possible by malicious actors with access to design information from the IP owner. Such chips are not accounted for by the IP owner and will not be enrolled under the IP owner's account and ergo cannot be sold under the original brand. The support for active hardware metering within the framework ensures that such chips are not functional even if enrolled under a different brand.

\subsubsection{Remarking and illegitimate-recycling}
Often bootleggers try to tamper with the markings on the package of an IC to sell it under the guise of a higher-grade product. Our framework stores all information present on the package of the IC and can immediately catch such discrepancies and invalidate a transaction. A similar situation is presented when bootleggers try to sell a recycled product under the guise of being brand new, which is easily caught with the production information.

\subsubsection{Defective, Out-of-spec, Forged certification and Tampering}
The framework allows for owners to report malfunctioning ICs and ensures that such devices do not circulate in the supply chain.

\subsection{Comparative Analysis}

In table \ref{tab:comparisons} we compare the ICtoken framework with prior works leveraging blockchain technology to monitor the IC supply chain. \cite{Islam19}, \cite{Chaudhary21} and \cite{Cui19} do not track the production stage or status of an IC and can not provide product-level provenance. \cite{Cui19}, \cite{Xu19} and  \cite{Vosatka20} propose consortium managed frameworks that are not transparent. They only permit entities within the consortium to utilize the system. Except for \cite{Cui19} all of the other works store details of genuine ICs in the plain and do not account for data privacy. \cite{Islam19}, \cite{Xu19} and \cite{Vosatka20} store CRPs for authenticating an IC leading to huge transaction sizes. Unlike all previous works, ICtoken can securely store active metering information for an IC in the form of an in-borne \textit{public key} which can then be used to validate its authenticity.

\begin{table}[h!]
    \centering
    \begin{tabular}{|c|cccccc|}
    \hline
          \textbf{Framework} & \textbf{OT} & \textbf{PT} & \textbf{Tr} & \textbf{DP}   & \textbf{AM} & \textbf{TS} \\
\hline
\cite{Islam19}    &\cmark&\xmark&\cmark&\xmark&\xmark& at least 2KB \\
\hline
\cite{Chaudhary21}&\cmark&\xmark&\cmark&\xmark&\xmark & - \\
\hline
\cite{Cui19}      &\cmark&\xmark&\xmark& \cmark &\xmark& - \\
\hline
\cite{Xu19}       &\cmark&\cmark&\xmark& \xmark &\xmark& at least 2KB \\
\hline
\cite{Vosatka20}  &\cmark&\cmark&\xmark&\xmark&\xmark& at least 2KB \\
\hline
\textbf{ICtoken}           &\cmark&\cmark&\cmark&\cmark&\cmark& \textbf{0.7KB}  \\
\hline
\end{tabular}
    
\caption{Comparison with prior works. [OT: Ownership Tracking; PT: Production Tracking; Tr: Transparency; DP: Data Privacy; AM: Active Metering; TS:Transaction Size]}
\label{tab:comparisons}
\end{table}

\section{Conclusion}

In this paper, we address the supply chain threats to the IP owner that arise due to the globalization of the electronics supply chain. The main problem with prior work in this area is that they are not designed with the end-product in mind and do not serve much purpose to the end-user. The proposed framework mitigates their shortcomings to offer enhanced IP integrity and stronger  assurance of a genuine product to the end-users. The main contribution of this work is the idea of ICtoken as a building block NFT unit for a product-level NFT. The transparent design with certified metadata of all components used in the product fortifies credibility across the supply chain and authenticity of end-product.

\bibliographystyle{ACM-Reference-Format}
\bibliography{ref}


\begin{thebibliography}{39}


\ifx \showCODEN    \undefined \def \showCODEN     #1{\unskip}     \fi
\ifx \showDOI      \undefined \def \showDOI       #1{#1}\fi
\ifx \showISBNx    \undefined \def \showISBNx     #1{\unskip}     \fi
\ifx \showISBNxiii \undefined \def \showISBNxiii  #1{\unskip}     \fi
\ifx \showISSN     \undefined \def \showISSN      #1{\unskip}     \fi
\ifx \showLCCN     \undefined \def \showLCCN      #1{\unskip}     \fi
\ifx \shownote     \undefined \def \shownote      #1{#1}          \fi
\ifx \showarticletitle \undefined \def \showarticletitle #1{#1}   \fi
\ifx \showURL      \undefined \def \showURL       {\relax}        \fi
\providecommand\bibfield[2]{#2}
\providecommand\bibinfo[2]{#2}
\providecommand\natexlab[1]{#1}
\providecommand\showeprint[2][]{arXiv:#2}

\bibitem[Anonymous(2022)]%
        {implementations}
\bibfield{author}{\bibinfo{person}{Anonymous}.} \bibinfo{year}{2022}\natexlab{}.
\newblock \bibinfo{title}{ICtoken Implementations}.
\newblock
\newblock
\newblock
\shownote{This citation has been redacted to protect the anonymity of the source}.


\bibitem[Barry(2021)]%
        {radicalSC}
\bibfield{author}{\bibinfo{person}{Quinn Barry}.} \bibinfo{year}{2021}\natexlab{}.
\newblock \bibinfo{title}{Inside Radical Semiconductor: The Stanford Startup Disrupting the Chip Industry}.
\newblock \bibinfo{howpublished}{Identity Review}.
\newblock
\urldef\tempurl%
\url{https://identityreview.com/inside-radical-semiconductor-the-stanford-startup-disrupting-the-chip-industry/}
\showURL{%
\tempurl}


\bibitem[Beerel et~al\mbox{.}(2022)]%
        {Beerel22}
\bibfield{author}{\bibinfo{person}{Peter Beerel}, \bibinfo{person}{Marios Georgiou}, \bibinfo{person}{Ben Hamlin}, \bibinfo{person}{Alex~J. Malozemoff}, {and} \bibinfo{person}{Pierluigi Nuzzo}.} \bibinfo{year}{2022}\natexlab{}.
\newblock \bibinfo{title}{Towards a Formal Treatment of Logic Locking}.
\newblock \bibinfo{howpublished}{Cryptology ePrint Archive, Report 2022/503}.
\newblock
\newblock
\shownote{\url{https://ia.cr/2022/503}}.


\bibitem[Benet(2014)]%
        {Benet14}
\bibfield{author}{\bibinfo{person}{Juan Benet}.} \bibinfo{year}{2014}\natexlab{}.
\newblock \bibinfo{title}{IPFS - Content Addressed, Versioned, P2P File System}.
\newblock \bibinfo{howpublished}{IPFS Whitepaper}.
\newblock
\urldef\tempurl%
\url{https://ipfs.io/ipfs/QmR7GSQM93Cx5eAg6a6yRzNde1FQv7uL6X1o4k7zrJa3LX/ipfs.draft3.pdf}
\showURL{%
\tempurl}


\bibitem[Buterin(2014)]%
        {Buterin14}
\bibfield{author}{\bibinfo{person}{Vitalik Buterin}.} \bibinfo{year}{2014}\natexlab{}.
\newblock \bibinfo{title}{Ethereum: A Next-Generation Smart Contract and Decentralized Application Platform}.
\newblock \bibinfo{howpublished}{Ethereum Whitepaper}.
\newblock
\urldef\tempurl%
\url{https://ethereum.org/en/whitepaper/}
\showURL{%
\tempurl}


\bibitem[Chaudhary et~al\mbox{.}(2021)]%
        {Chaudhary21}
\bibfield{author}{\bibinfo{person}{Chandan~Kumar Chaudhary}, \bibinfo{person}{Urbi Chatterjee}, {and} \bibinfo{person}{Debdeep Mukhopadhayay}.} \bibinfo{year}{2021}\natexlab{}.
\newblock \bibinfo{title}{Auto-PUFChain: An Automated Interaction Tool for PUFs and Blockchain in Electronic Supply Chain}.
\newblock , \bibinfo{numpages}{4}~pages.
\newblock
\urldef\tempurl%
\url{https://doi.org/10.1109/AsianHOST53231.2021.9699720}
\showDOI{\tempurl}


\bibitem[Chhotaray and Shrimpton(2021)]%
        {Chhotaray21}
\bibfield{author}{\bibinfo{person}{Animesh Chhotaray} {and} \bibinfo{person}{Thomas Shrimpton}.} \bibinfo{year}{2021}\natexlab{}.
\newblock \bibinfo{title}{Hardening Circuit-Design IP Against Reverse-Engineering Attacks}.
\newblock \bibinfo{howpublished}{Cryptology ePrint Archive, Report 2021/456}.
\newblock
\newblock
\shownote{\url{https://ia.cr/2021/456}}.


\bibitem[Contreras et~al\mbox{.}(2013)]%
        {Contreras13}
\bibfield{author}{\bibinfo{person}{Gustavo~K. Contreras}, \bibinfo{person}{Md.~Tauhidur Rahman}, {and} \bibinfo{person}{Mohammad Tehranipoor}.} \bibinfo{year}{2013}\natexlab{}.
\newblock \bibinfo{title}{Secure Split-Test for preventing IC piracy by untrusted foundry and assembly}.
\newblock , \bibinfo{numpages}{196-203}~pages.
\newblock
\urldef\tempurl%
\url{https://doi.org/10.1109/DFT.2013.6653606}
\showDOI{\tempurl}


\bibitem[Counterintelligence and Center(2022)]%
        {scsupshainrisks}
\bibfield{author}{\bibinfo{person}{National Counterintelligence} {and} \bibinfo{person}{Security Center}.} \bibinfo{year}{2022}\natexlab{}.
\newblock \bibinfo{title}{Supply Chain Risks to Semiconductors}.
\newblock
\newblock
\urldef\tempurl%
\url{https://www.dni.gov/files/NCSC/documents/supplychain/semiconductor-supply-chain-2022-39E2C6B0-.pdf}
\showURL{%
\tempurl}


\bibitem[Cui et~al\mbox{.}(2019)]%
        {Cui19}
\bibfield{author}{\bibinfo{person}{Pinchen Cui}, \bibinfo{person}{Julie Dixon}, \bibinfo{person}{Ujjwal Guin}, {and} \bibinfo{person}{Daniel Dimase}.} \bibinfo{year}{2019}\natexlab{}.
\newblock \bibinfo{title}{A Blockchain-Based Framework for Supply Chain Provenance}.
\newblock , \bibinfo{numpages}{157113-157125}~pages.
\newblock
\urldef\tempurl%
\url{https://doi.org/10.1109/ACCESS.2019.2949951}
\showDOI{\tempurl}


\bibitem[Entriken et~al\mbox{.}(2018)]%
        {eip721}
\bibfield{author}{\bibinfo{person}{William Entriken}, \bibinfo{person}{Dieter Shirley}, \bibinfo{person}{Jacob Evans}, {and} \bibinfo{person}{Nastassia Sachs}.} \bibinfo{year}{2018}\natexlab{}.
\newblock \bibinfo{title}{EIP-721: Non-Fungible Token Standard}.
\newblock \bibinfo{howpublished}{Ethereum Improvement Proposals}.
\newblock
\urldef\tempurl%
\url{https://eips.ethereum.org/EIPS/eip-721}
\showURL{%
\tempurl}


\bibitem[Ethereum(2018)]%
        {erc721}
\bibfield{author}{\bibinfo{person}{Ethereum}.} \bibinfo{year}{2018}\natexlab{}.
\newblock \bibinfo{title}{ERC-721 NON-FUNGIBLE TOKEN STANDARD}.
\newblock
\newblock
\urldef\tempurl%
\url{https://ethereum.org/en/developers/docs/standards/tokens/erc-721/}
\showURL{%
\tempurl}


\bibitem[Ethereum(2022)]%
        {solidity}
\bibfield{author}{\bibinfo{person}{Ethereum}.} \bibinfo{year}{2022}\natexlab{}.
\newblock \bibinfo{title}{Solidity Documentation Release 0.8.15}.
\newblock
\newblock
\urldef\tempurl%
\url{https://buildmedia.readthedocs.org/media/pdf/solidity/develop/solidity.pdf}
\showURL{%
\tempurl}


\bibitem[Ganji et~al\mbox{.}(2019)]%
        {Ganji19}
\bibfield{author}{\bibinfo{person}{Fatemeh Ganji}, \bibinfo{person}{Shahin Tajik}, \bibinfo{person}{Jean-Pierre Seifert}, {and} \bibinfo{person}{Domenic Forte}.} \bibinfo{year}{2019}\natexlab{}.
\newblock \bibinfo{title}{Blockchain-enabled Cryptographically-secure Hardware Obfuscation}.
\newblock \bibinfo{howpublished}{Cryptology ePrint Archive, Report 2019/928}.
\newblock
\newblock
\shownote{\url{https://ia.cr/2019/928}}.


\bibitem[Guin et~al\mbox{.}(2014)]%
        {guin2014detectandprevent}
\bibfield{author}{\bibinfo{person}{Ujjwal Guin}, \bibinfo{person}{Daniel DiMase}, {and} \bibinfo{person}{Mohammad Tehranipoor}.} \bibinfo{year}{2014}\natexlab{}.
\newblock \bibinfo{title}{Counterfeit integrated circuits: Detection, avoidance, and the challenges ahead}.
\newblock , \bibinfo{numpages}{9--23}~pages.
\newblock
\urldef\tempurl%
\url{https://doi.org/10.1007/s10836-013-5430-8}
\showURL{%
\tempurl}


\bibitem[Herder et~al\mbox{.}(2014)]%
        {herder14}
\bibfield{author}{\bibinfo{person}{Charles Herder}, \bibinfo{person}{Meng-Day Yu}, \bibinfo{person}{Farinaz Koushanfar}, {and} \bibinfo{person}{Srinivas Devadas}.} \bibinfo{year}{2014}\natexlab{}.
\newblock \bibinfo{title}{Physical Unclonable Functions and Applications: A Tutorial}.
\newblock , \bibinfo{numpages}{1126-1141}~pages.
\newblock
\urldef\tempurl%
\url{https://doi.org/10.1109/JPROC.2014.2320516}
\showDOI{\tempurl}


\bibitem[ID(2017)]%
        {intrinsic}
\bibfield{author}{\bibinfo{person}{INTRINSIC ID}.} \bibinfo{year}{2017}\natexlab{}.
\newblock \bibinfo{title}{SRAM PUF : The Secure Silicon Fingerprint}.
\newblock
\newblock
\newblock
\shownote{\url{https://www.intrinsic-id.com/resources/white-papers/white-paper-sram-puf-secure-silicon-fingerprint/}}.


\bibitem[Islam and Kundu(2019)]%
        {Islam19}
\bibfield{author}{\bibinfo{person}{Md~Nazmul Islam} {and} \bibinfo{person}{Sandip Kundu}.} \bibinfo{year}{2019}\natexlab{}.
\newblock \bibinfo{title}{Enabling IC Traceability via Blockchain Pegged to Embedded PUF}.
\newblock , \bibinfo{numpages}{23}~pages.
\newblock
\showISSN{1084-4309}
\urldef\tempurl%
\url{https://doi.org/10.1145/3315669}
\showDOI{\tempurl}


\bibitem[Kamali et~al\mbox{.}(2022)]%
        {kamali22}
\bibfield{author}{\bibinfo{person}{Hadi~Mardani Kamali}, \bibinfo{person}{Kimia~Zamiri Azar}, \bibinfo{person}{Farimah Farahmandi}, {and} \bibinfo{person}{Mark Tehranipoor}.} \bibinfo{year}{2022}\natexlab{}.
\newblock \bibinfo{title}{Advances in Logic Locking: Past, Present, and Prospects}.
\newblock \bibinfo{howpublished}{Cryptology ePrint Archive, Report 2022/260}.
\newblock
\newblock
\shownote{\url{https://ia.cr/2022/260}}.


\bibitem[Koushanfar(2010)]%
        {Koushanfar10}
\bibfield{author}{\bibinfo{person}{Farinaz Koushanfar}.} \bibinfo{year}{2010}\natexlab{}.
\newblock \bibinfo{title}{Hardware Metering: A Survey}.
\newblock
\newblock
\showISBNx{978-1-4419-8079-3}
\urldef\tempurl%
\url{https://doi.org/10.1007/978-1-4419-8080-9_5}
\showDOI{\tempurl}


\bibitem[Leef(2018)]%
        {Serge18}
\bibfield{author}{\bibinfo{person}{Serge Leef}.} \bibinfo{year}{2018}\natexlab{}.
\newblock \bibinfo{title}{Supply Chain Hardware Integrity for Electronics Defense}.
\newblock
\newblock
\urldef\tempurl%
\url{https://csrc.nist.gov/CSRC/media/Projects/cyber-supply-chain-risk-management/documents/SSCA/Winter_2018/TuePM2.1-SHIELD.pdf}
\showURL{%
\tempurl}


\bibitem[Monax(2019)]%
        {monax}
\bibfield{author}{\bibinfo{person}{Monax}.} \bibinfo{year}{2019}\natexlab{}.
\newblock
\newblock
\urldef\tempurl%
\url{https://monax.io/help/advanced/nft-integration/}
\showURL{%
\tempurl}


\bibitem[Nakamoto(2008)]%
        {Nakamoto08}
\bibfield{author}{\bibinfo{person}{Satoshi Nakamoto}.} \bibinfo{year}{2008}\natexlab{}.
\newblock \bibinfo{title}{Bitcoin: A Peer-to-Peer Electronic Cash System}.
\newblock \bibinfo{howpublished}{Bitcoin Whitepaper}.
\newblock
\urldef\tempurl%
\url{https://bitcoin.org/en/bitcoin-paper}
\showURL{%
\tempurl}


\bibitem[Nehra(2021)]%
        {Nehra21}
\bibfield{author}{\bibinfo{person}{Mahipal Nehra}.} \bibinfo{year}{2021}\natexlab{}.
\newblock \bibinfo{title}{Blockchain in Supply Chain: A Transparent Prospect for Products}.
\newblock \bibinfo{howpublished}{ReadWrite}.
\newblock
\urldef\tempurl%
\url{https://readwrite.com/blockchain-in-supply-chain-a-transparent-prospect-for-products/}
\showURL{%
\tempurl}


\bibitem[Newton(2022)]%
        {Newton22}
\bibfield{author}{\bibinfo{person}{Emily Newton}.} \bibinfo{year}{2022}\natexlab{}.
\newblock \bibinfo{title}{Is the chip shortage leading to more counterfeit components?}
\newblock \bibinfo{howpublished}{Embedded.com}.
\newblock
\urldef\tempurl%
\url{https://www.embedded.com/is-the-chip-shortage-leading-to-more-counterfeit-components/}
\showURL{%
\tempurl}


\bibitem[of~Commerce: Bureau~of Industry and of~Technology~Evaluation(2010)]%
        {usdoc}
\bibfield{author}{\bibinfo{person}{U.S.~Department of~Commerce: Bureau~of Industry} {and} \bibinfo{person}{Security~Office of Technology~Evaluation}.} \bibinfo{year}{2010}\natexlab{}.
\newblock \bibinfo{title}{Defense Industrial Base Assessment: Counterfeit Electronics}.
\newblock
\newblock
\urldef\tempurl%
\url{https://agmaglobal.org/uploads/BIS\%20Survey\%20(January\%202010\%20final).pdf}
\showURL{%
\tempurl}


\bibitem[Office(2016)]%
        {goa16}
\bibfield{author}{\bibinfo{person}{United States Government~Accountability Office}.} \bibinfo{year}{2016}\natexlab{}.
\newblock \bibinfo{title}{Counterfeit Parts: DOD Needs to Improve Reporting and Oversight to Reduce Supply Chain Risk}.
\newblock
\newblock
\urldef\tempurl%
\url{https://www.gao.gov/assets/gao-16-236.pdf}
\showURL{%
\tempurl}


\bibitem[Rajendran(2017)]%
        {Rajendran17}
\bibfield{author}{\bibinfo{person}{Jeyavijayan J~V Rajendran}.} \bibinfo{year}{2017}\natexlab{}.
\newblock \bibinfo{title}{An overview of hardware intellectual property protection}.
\newblock , \bibinfo{numpages}{4}~pages.
\newblock
\urldef\tempurl%
\url{https://doi.org/10.1109/ISCAS.2017.8050883}
\showDOI{\tempurl}


\bibitem[Rostami et~al\mbox{.}(2014)]%
        {Rostami14}
\bibfield{author}{\bibinfo{person}{Masoud Rostami}, \bibinfo{person}{Farinaz Koushanfar}, {and} \bibinfo{person}{Ramesh Karri}.} \bibinfo{year}{2014}\natexlab{}.
\newblock \bibinfo{title}{A Primer on Hardware Security: Models, Methods, and Metrics}.
\newblock , \bibinfo{numpages}{1283-1295}~pages.
\newblock
\urldef\tempurl%
\url{https://doi.org/10.1109/JPROC.2014.2335155}
\showDOI{\tempurl}


\bibitem[Roy et~al\mbox{.}(2008)]%
        {Roy08}
\bibfield{author}{\bibinfo{person}{Jarrod~A. Roy}, \bibinfo{person}{Farinaz Koushanfar}, {and} \bibinfo{person}{Igor~L. Markov}.} \bibinfo{year}{2008}\natexlab{}.
\newblock \bibinfo{title}{EPIC: Ending Piracy of Integrated Circuits}.
\newblock , \bibinfo{numpages}{1069-1074}~pages.
\newblock
\urldef\tempurl%
\url{https://doi.org/10.1109/DATE.2008.4484823}
\showDOI{\tempurl}


\bibitem[skuchainweb(2016)]%
        {skuchain}
\bibfield{author}{\bibinfo{person}{skuchainweb}.} \bibinfo{year}{2016}\natexlab{}.
\newblock \bibinfo{title}{Skuchain: Here’s how blockchain will save global trade a trillion dollars}.
\newblock
\newblock
\urldef\tempurl%
\url{https://www.skuchain.com/skuchain-heres-how-blockchain-will-save-global-trade-a-trillion-dollars/}
\showURL{%
\tempurl}


\bibitem[Software(2019)]%
        {fedex}
\bibfield{author}{\bibinfo{person}{Precision Software}.} \bibinfo{year}{2019}\natexlab{}.
\newblock \bibinfo{title}{Could Blockchain Revolutionize Parcel Shipping?}
\newblock
\newblock
\urldef\tempurl%
\url{https://www.fedex.com/content/dam/fedex/us-united-states/Compatible-Solutions/images/2019/Q2/Could_Blockchain_Revolutionize_Parcel_Shipping_V2_50457811.pdf}
\showURL{%
\tempurl}


\bibitem[Sristy(2021)]%
        {walmartFood}
\bibfield{author}{\bibinfo{person}{Archana Sristy}.} \bibinfo{year}{2021}\natexlab{}.
\newblock \bibinfo{title}{National Counterintelligence and Security Center}.
\newblock
\newblock
\urldef\tempurl%
\url{https://one.walmart.com/content/globaltechindia/en_in/Tech-insights/blog/Blockchain-in-the-food-supply-chain.html}
\showURL{%
\tempurl}


\bibitem[the~largest NFT~marketplace(2017)]%
        {opensea}
\bibfield{author}{\bibinfo{person}{OpenSea the~largest NFT~marketplace}.} \bibinfo{year}{2017}\natexlab{}.
\newblock
\newblock
\urldef\tempurl%
\url{https://opensea.io/}
\showURL{%
\tempurl}


\bibitem[Vosatka et~al\mbox{.}(2020)]%
        {Vosatka20}
\bibfield{author}{\bibinfo{person}{Jason Vosatka}, \bibinfo{person}{Andrew Stern}, \bibinfo{person}{M.M. Hossain}, \bibinfo{person}{Fahim Rahman}, \bibinfo{person}{Jeffery Allen}, \bibinfo{person}{Monica Allen}, \bibinfo{person}{Farimah Farahmandi}, {and} \bibinfo{person}{Mark Tehranipoor}.} \bibinfo{year}{2020}\natexlab{}.
\newblock \bibinfo{title}{Tracking Cloned Electronic Components using a Consortium-based Blockchain Infrastructure}.
\newblock , \bibinfo{numpages}{6}~pages.
\newblock
\urldef\tempurl%
\url{https://doi.org/10.1109/PAINE49178.2020.9337735}
\showDOI{\tempurl}


\bibitem[Wang et~al\mbox{.}(2021)]%
        {Wang21}
\bibfield{author}{\bibinfo{person}{Qin Wang}, \bibinfo{person}{Rujia Li}, \bibinfo{person}{Qi Wang}, {and} \bibinfo{person}{Shiping Chen}.} \bibinfo{year}{2021}\natexlab{}.
\newblock \bibinfo{title}{Non-Fungible Token (NFT): Overview, Evaluation, Opportunities and Challenges}.
\newblock
\newblock


\bibitem[Xu et~al\mbox{.}(2019)]%
        {Xu19}
\bibfield{author}{\bibinfo{person}{Xiaolin Xu}, \bibinfo{person}{Fahim Rahman}, \bibinfo{person}{Bicky Shakya}, \bibinfo{person}{Apostol Vassilev}, \bibinfo{person}{Domenic Forte}, {and} \bibinfo{person}{Mark Tehranipoor}.} \bibinfo{year}{2019}\natexlab{}.
\newblock \bibinfo{title}{Electronics Supply Chain Integrity Enabled by Blockchain}.
\newblock , \bibinfo{numpages}{25}~pages.
\newblock
\showISSN{1084-4309}
\urldef\tempurl%
\url{https://doi.org/10.1145/3315571}
\showDOI{\tempurl}


\bibitem[Yasin and Sinanoglu(2017)]%
        {Yasin17}
\bibfield{author}{\bibinfo{person}{Muhammad Yasin} {and} \bibinfo{person}{Ozgur Sinanoglu}.} \bibinfo{year}{2017}\natexlab{}.
\newblock \bibinfo{title}{Evolution of Logic Locking}.
\newblock
\newblock
\urldef\tempurl%
\url{https://doi.org/10.1109/VLSI-SoC.2017.8203496}
\showDOI{\tempurl}


\bibitem[Young(2022)]%
        {EY22}
\bibfield{author}{\bibinfo{person}{Ernst~\& Young}.} \bibinfo{year}{2022}\natexlab{}.
\newblock \bibinfo{title}{EY OpsChain Contract Manager}.
\newblock \bibinfo{howpublished}{ey.com}.
\newblock
\urldef\tempurl%
\url{https://www.ey.com/en_gl/blockchain-platforms/contract-manager}
\showURL{%
\tempurl}


\end{thebibliography}

\appendix

\end{document}